\documentclass[11pt]{article}

\title{Status of the Standard Model}

\author{Paul H. Frampton\\
University of North Carolina at Chapel Hill}

\date{October 2003}

\begin{document}
\maketitle

\section{Introduction}
 
The standard model is healthy is all respects
except for the non-zero neutrino masses which
require an extension of the minimal version.

This talk was in three parts:

(I) Rise and Fall of the Zee Model 1998-2001.

(II) Classification of Two-Zero Textures.

(III) FGY Model relating Cosmological B to Neutrino CP violation.

For parts (I) and (II) references are provided. Part (III) is
included in this write-up.

\section{Rise and Fall of the Zee Model 1998-2001}

This first part was based on:

P.H. Frampton and S.L. Glashow, Phys. Lett.
{\bf 461,} 95 (1999). {\tt hep-ph/9906375}.

\section{Classification of Two-Zero Textures}

The second part was based on:

P.H. Frampton, S.L. Glashow and D. Marfatia,
Phys. Lett. {\bf B536,} 79 (2002).
{\tt hep-ph/0201008}

\section{FGY Model relating Cosmological B with Neutrino CP Violation}

One of the most profound ideas is\cite{Sakharov}
that baryon number asymmetry arises in the early universe
because of processes which violate CP symmetry and that
terrestrial experiments on CP violation
could therefore inform us of the details of such
cosmological baryogenesis.

The early discussions of baryogenesis focused on
the violation of baryon number and its possible relation to
proton decay. In the light of present evidence for neutrino masses
and oscillations
it is more fruitful to associate the baryon number
of the universe with violation of lepton number\cite{FY}.
In the present Letter
we shall show how, in one class of models, the sign of the
baryon number of the universe correlates with
the results of CP violation in neutrino oscillation
experiments which will be performed in the forseeable
future.

Present data on atmospheric and solar neutrinos suggest
that there are respective squared mass differences
$\Delta_a \simeq 3 \times 10^{-3} eV^2$
and
$\Delta_s \simeq 5 \times 10^{-5} eV^2$.
The corresponding mixing angles $\theta_1$
and $\theta_3$ satisfy
$tan^2 \theta_1 \simeq 1$
and $0.6 \leq sin^2 2\theta_3 \leq 0.96$
with $sin^2 \theta_3 = 0.8$ as the best fit.
The third mixing angle is much smaller than the other
two, since the data require $sin^2 2 \theta_2 \leq 0.1$.

A first requirement is that our model\cite{FGY} accommodate
these experimental facts at low energy.

\bigskip
\bigskip

\section{The Model}

\bigskip
\bigskip

In the minimal standard model, neutrinos are massless.
The most economical addition to the standard model
which accommodates both neutrino masses
and allows the violation
of lepton number to underly the cosmological baryon asymmetry
is two right-handed neutrinos $N_{1,2}$.

These lead to new terms in the lagrangian:

\begin{eqnarray}
{\cal L} & = & \frac{1}{2} (N_1, N_2) \left(
\begin{array}{cc} M_1 & 0 \\ 0 & M_2
\end{array} \right)
\left( \begin{array}{c} N_1 \\ N_2 \end{array} \right)
+  \nonumber \\
& + & (N_1, N_2)
\left( \begin{array}{ccc} a  &  a^{'}  &  0  \\
0  &  b  &  b^{'}  \end{array} \right)
\left( \begin{array}{c} l_1  \\  l_2 \\ l_3 \end{array}
\right)  H  + h.c.
\label{Lag}
\end{eqnarray}
where we shall denote the rectangular Dirac mass matrix
by $D_{ij}$. We have assumed a texture
for $D_{ij}$ in which the upper
right and lower left entries vanish.
The remaining parameters in our model
are both necessary and sufficient
to account for the data.

For the light neutrinos, the see-saw mechanism leads to
the mass matrix\cite{Y}
\begin{eqnarray}
\hat{L} & = & D^T M^{-1} D \nonumber \\
& = & \left( \begin{array}{ccc}
\frac{a^2}{M_1}  &  \frac{a a^{'}}{M_1}  &  0  \\
\frac{a a^{'}}{M_1}  &  \frac{(a^{'})^2}{M_1} + \frac{b^2}{M_2}
&  \frac{b b^{'}}{M_2} \\
0  &  \frac{b b^{'}}{M_2}  &  \frac{(b^{'})^2}{M_2}
\end{array}  \right)
\label{L}
\end{eqnarray}

We take a basis where $a, b, b^{'}$ are real and where
$a^{'}$ is complex $a^{'} \equiv |a^{'}|e^{i \delta}$.
To check consistency with low-energy
phenomenology we temporarily take the specific
values (these will be loosened later)
$b^{'} = b$ and $a^{'} = \sqrt{2} a$ and all parameters real.
In that case:
\begin{eqnarray}
\hat{L}
& = & \left( \begin{array}{ccc}
\frac{a^2}{M_1}  &  \frac{\sqrt{2}a^2}{M_1}  &  0  \\
\frac{\sqrt{2}a^2}{M_1}  &  \frac{2a^2}{M_1} + \frac{b^2}{M_2}
&  \frac{b^{2}}{M_2} \\
0  &  \frac{b^{2}}{M_2}  &  \frac{b^{^2}}{M_2}
\end{array}  \right)
\label{LL}
\end{eqnarray}
We now diagonalize to the mass basis by writing:
\begin{equation}
{\cal L} = \frac{1}{2} \nu^T \hat{L} \nu
= \frac{1}{2} \nu^{'T} U^T \hat{L} U \nu^{'}
\end{equation}
where
\begin{eqnarray}
U & = & \left( \begin{array}{ccc} 1/\sqrt{2}  &  1/\sqrt{2}  &  0  \\
- 1/2 & 1/2  &  1/\sqrt{2}  \\
1/2  &  -1/2 &  1/\sqrt{2}
\end{array} \right) \times \nonumber \\
& \times &
\left( \begin{array}{ccc}
1  &  0  &  0  \\
0  &  cos\theta &  sin\theta  \\
0  &  - sin\theta  &  cos\theta
\end{array}
\right)
\end{eqnarray}
We deduce that the mass eigenvalues  and $\theta$
are given by

\begin{equation}
m(\nu_3^{'}) \simeq 2 b^2/M_2; ~~~ m(\nu_2^{'}) \simeq 2 a^2 /M_1; ~
~~
m(\nu_1^{'}) = 0
\end{equation}
and
\begin{equation}
\theta \simeq m(\nu_2^{'}) / (\sqrt{2} m(\nu_3^{'}))
\end{equation}
in which it was assumed that $a^2/M_1 \ll b^2/M_2$.

By examining the relation between the three mass eigenstates
and the corresponding flavor eigenstates
we find
that for the unitary matrix relevant to neutrino oscillations
that
\begin{equation}
U_{e3} \simeq sin\theta/\sqrt{2} \simeq m(\nu_2)/(2m(\nu_3))
\end{equation}

Thus the assumptions $a^{'} = \sqrt{2} a$, $
b^{'} = b$ adequately fit the
experimental data, but
$a^{'}$ and $b^{'}$ could
be varied around
$\sqrt{2}a$ and $b$ respectively
to achieve better fits.

But we may conclude that
\begin{eqnarray}
2b^2/M_2 & \simeq & 0.05 eV = \sqrt{\Delta_a} \nonumber \\
2a^2/M_1 & \simeq & 7 \times 10^{-3} eV  = \sqrt{\Delta_{s}}
\label{numass}
\end{eqnarray}

It follows from these values that $N_1$ decay satisfies the
out-of-equilibrium condition for leptogenesis (the
absolute requirement is $m < 10^{-2} eV$ \cite{buchpascos})
while $N_2$ decay does not.
This fact enables us to predict
the sign of CP violation in neutrino oscillations without ambiguity.

\bigskip
\bigskip

\section{Connecting Link}

\bigskip
\bigskip

Let us now come to the main result. Having
a model consistent with all low-energy data and with
adequate texture zeros\cite{FGM} in $\hat{L}$ and
equivalently $D$ we can
compute the sign both of the high-energy
CP violating parameter ($\xi_H$) appearing
in leptogenesis and of the CP violation parameter
which will be measured in low-energy
$\nu$ oscillations ($\xi_L$).

We find the baryon number $B$ of the universe
produced by $N_1$ decay
proportional to\cite{Buch}
\begin{eqnarray}
B & \propto & \xi_H =
(Im D D^{\dagger} )_{12}^2 = Im (a^{'} b)^2 \nonumber \\
& = & + Y^2a^2b^2 sin 2\delta
\label{highenergy}
\end{eqnarray}
in which $B$ is positive by observation of the universe.
Here we have loosened our assumption about $a^{'}$
to $a^{'} = Y a e^{i \delta}$.

At low energy the CP violation in neutrino oscillations is governed
by
the quantity\cite{branco}
\begin{equation}
\xi_L = Im (h_{12} h_{23} h_{31})
\end{equation}
where $h = \hat{L} \hat{L}^{\dagger}$.

Using Eq.(\ref{L}) we find:
\begin{eqnarray}
h_{12} & = & \left( \frac{a^3 a^{'*}}{M_1^2} + \frac{a |a^{'}|^2 a^{
'*}}{M_1^2}
\right) + \frac{a a^{'} b^2}{M_1M_2} \nonumber \\
h_{23} & = & \left( \frac{b b^{'} a^{'2}}{M_1 M_2} \right)
 + \left( \frac{b^3 b^{'}}{M_2^2} + \frac{b b^{'3}}{M_2^2} \right)
\nonumber \\
h_{31} & = & \left( \frac{a a^{'*} b b^{'}}{M_1 M_2}
\right) \nonumber \\
\end{eqnarray}
from which it follows that
\begin{equation}
\xi_L =  - \frac{a^6 b^6}{M_1^3 M_2^3} sin 2\delta [ Y^2 (2 + Y^2)]
\label{lowenergy}
\end{equation}
Here we have taken $b=b^{'}$ because
the mixing for the atmospheric neutrinos
is almost maximal.

Neutrinoless double beta decay $(\beta\beta)_{0\nu}$
is predicted at a rate corresponding to $\hat{L}_{ee} \simeq 3 \times 10^{-3}eV$.

The comparison between Eq.(\ref{highenergy})
and Eq.(\ref{lowenergy})
now gives a unique relation between the signs of $\xi_L$ and $\xi_H$
.

As a check of this assertion we consider
the equally viable alternative model

\begin{equation}
D = \left( \begin{array}{ccc} a & 0 & a^{'} \\
0 & b & b^{'}
\end{array} \right)
\label{alternative}
\end{equation}
in Eq.(\ref{Lag})
where $\xi_L$ reverses sign but the
signs of $\xi_H$ and $\xi_L$ are
still uniquely correlated once the $\hat{L}$
textures arising from the $D$ textures
of Eq.(\ref{Lag}) and Eq.(\ref{alternative})
are distinguished by low-energy phenomenology.
Note that such models have five parameters including
a phase
and that cases B1 and B2 in \cite{FGM}
can be regarded as (unphysical) limits of (\ref{Lag})
and (\ref{alternative}) respectively.

This fulfils in such a class of models the idea of \cite{Sakharov}
with only the small change that baryon
number violation is replaced by lepton number violation.

\bigskip
\bigskip

\section{Further Properties}

\bigskip
\bigskip

The model of \cite{FGY} has additional properties
which we allude to here briefly:

\bigskip

\noindent 1) It is important that the zeroes occurring
in Eq.(\ref{Lag}) can be associated with a global symmetry and hence
 are
not infinitely renormalized. This can be achieved.

\bigskip
\bigskip

\noindent 2) The model has four parameters in the texture
of Eq.(\ref{L})  and leads to a prediction of
$\theta_{13}$ in terms of the other four parameters
$\Delta_a, \Delta_S, \theta_{12},$ and $\theta_{23}$.
The result is that $\theta_{13}$ is predicted to be non-zero
with magnitude related to the smallness of
$\Delta_S/\Delta_a$.

\bigskip

\noindent Details of these properties are currently under
further investigation.

\bigskip
\bigskip

\section*{Acknowledgement}

\bigskip
\bigskip
I thank  Professor N. Mankoc-Borstnik of the University
of Ljubjana
for organizing this stimulating workshop in Portoroz,
Slovenia.
This work was supported in part
by the Department of Energy
under Grant Number
DE-FG02-97ER-410236.

\end{document}